\documentclass[fleqn,usenatbib]{mnras}

\usepackage{newtxtext,newtxmath}

\usepackage[T1]{fontenc}
\usepackage{ae,aecompl}

\usepackage{amsmath}	
\usepackage{amssymb}	

\usepackage{ulem}  
\usepackage{color} 

\usepackage{fixltx2e}
\usepackage[flushleft]{threeparttable}

\usepackage{graphicx} 
\graphicspath{{figs/}}
\usepackage{rotating} 
\usepackage{caption}
\usepackage{mhchem} 

\newcommand\Msun  {{\rm M}_\odot}

\newcommand\etal  {{\it et al.\ }}
\newcommand\kms{\rm \, km ~ s^{-1}}
\newcommand\kmspc{\rm \, km ~ s^{-1} pc^{-1}}

\newcommand{\ppcc}{{\rm cm}^{-3}}

\newcommand\beq{\begin{equation}}
\newcommand\eeq{\end{equation}}

\title[Protostellar angular momentum] {On estimating angular momenta of infalling protostellar cores from observations}
\author[Zhang et al]
      { Shangjia Zhang$^{1}$,
      Lee Hartmann$^{1}$
       \thanks{E-mail:
          lhartm@umich.edu},
      Manuel Zamora-Avil\'es$^{1,2}$, and
      \newauthor\  Aleksandra Kuznetsova$^{1}$ %
\\
$^{1}$Department of Astronomy, University of Michigan, 1085 S.
University Ave, Ann Arbor, MI
48109 USA \\
$^{2}$Instituto de Radioastronom\'ia y Astrof\'isica,
            Universidad Nacional Aut\'onoma de M\'exico,
            Apdo. Postal 72-3 (Xangari), Morelia,\\
            Michoc\'an 58089, M\'exico \\
}

\date{Accepted XXX. Received YYY; in original form ZZZ}

\pubyear{2018}

\begin{document}
\label{firstpage}
\pagerange{\pageref{firstpage}--\pageref{lastpage}}
\maketitle

\begin{abstract}
We use numerical simulations of molecular cloud formation in the colliding flow scenario to investigate 
the reliability of observational estimates of the angular momenta of early-state, low-mass protostellar cores.  We show that, with suitable corrections 
for projection factors, molecular line observations
of velocity gradients in NH$_3$ can be used to provide 
reasonable estimates of core angular momenta within 
a factor of two to three, with a few large underestimates due
to unfavorable viewing angles.  Our results differ
from previous investigations which suggested that
observations might overestimate true angular momenta
by as much as an order of magnitude; the difference
is probably due to the much smoother velocity field
on small scales in our simulations, which result
from allowing turbulence to decay and gravitational
infall to dominate. The results emphasize the
importance of understanding the nature of ``turbulent''
velocities, with implications for the formation
of protostellar disks during core collapse.

\end{abstract}




\section{Introduction} \label{sec:intro}

The angular momenta of collapsing protostellar clouds establish the properties of binary and multiple stars and the initial conditions of their protoplanetary disks. Recent observations have demonstrated that disks of appreciable size can form early, during the Class I and even the Class 0 protostellar phases (\citealt{tobin12b,ohashi14,aso15,yen15,oya16}; see \citealt{takakuwa17} for a review).
However, simulations of collapsing magnetized protostellar cores over the last few years have indicated that magnetic fields can in some circumstances remove so much angular momentum that large disks do not form \citep[see, e.g.][for reviews]{seifried13,li14}.
Ambipolar diffusion can play a crucial role in reducing the magnetic flux responsible for angular momentum transport outward \citep{masson16,vaytet18}, although
it is unclear whether it is sufficient to allow the formation of disks during the Class 0 phase \citep{dapp12}.  Alternatively,
substantial misalignment of the magnetic field relative to the net rotation axis, and/or diffusion of the field in a turbulent environment, can allow early disk formation \citep{joos12,joos13,santos-lima12,santos-lima13,seifried12,seifried13,krumholz13,lewis15,wurster16,wurster17}.
In view of the theoretical uncertainties, observational estimates of infalling protostellar core angular momenta (and their directions) can be useful.
Following the classic work by \cite{goodman93} on the rotation of protostellar cores, additional observational estimates of angular momenta have been made
\citep{caselli02} \citep[see the review by][]{belloche13}, with
added emphasis on the angular momenta of inner envelopes \citep{yen15} to help constrain disk formation.

However, numerical simulations by \cite{offner08} and \cite{dib10} suggest that observations can overestimate true angular momenta by as much as a factor of 10.  These large bias factors presumably result from the assumed complexity of the turbulent velocity field, which cancels out some of the angular momentum on small scales that is smoothed out on large scales with the usual observational procedures.  This interpretation is supported by the results of \citeauthor{offner08} indicating that the bias is smaller if the initial seeded turbulent velocity field is allowed to decay rather than be continuously driven.

The approach to developing initial conditions that we favor is one in which the turbulence is the result of cloud formation, followed by gravitational collapse \citep[ e.g.,][]{Heitsch+06,bp2011}. The supersonic velocity fields at the time of star (sink particle) formation are then the result (mostly) of overdensities in the cloud which drive motion gravitationally.  This arguably results in a smoother velocity field than in the case of continued driving of turbulence by some unspecified mechanism; in turn, this could mean that observational estimates of angular momenta could be more representative of true values.  To investigate this further,
in this paper we employ the results of numerical simulations for dense cores, post-process them with radiative transfer,
and use the same methods of analysis for
closer comparison with observational results.

\section{Numerical methods} \label{sec:numerics}

\subsection{Numerical simulations with FLASH} \label{subsec:simulations}

The sub-parsec cores analyzed in this work come from numerical simulations at a much bigger scale (256 pc) presented in \citet[][model labeled B3J]{ZA+17}. This simulation was carried out with the Eulerian adaptive mesh refinement code FLASH \citep[version 2.5;][]{flash} with the objective of studying the effect of magnetic fields in the formation and evolution of molecular clouds (MCs). In that work, the simulated clouds are assembled by two warm neutral streams (or inflows), which collide at the center of the numerical box. The collision compresses the warm gas triggering a phase transition due to the thermal instability \citep{Field65}, thus producing cold gas. Moreover, the converging flows through different dynamical instabilities inject naturally the turbulence that characterizes MCs \cite[see, e.g.,][]{Heitsch+06}, which in turn produce the overdensities (seeds of cores). As the cold layer continues accreting material from the inflows, it increases its mass and column density, allowing the formation of molecular gas, and eventually the entire cloud enters in a state of hierarchical collapse \citep[see, e.g.,][]{Hartmann+01}.

The simulation includes the relevant physical processes for cloud formation/evolution such as heating and cooling, self-gravity, magnetic fields, and sink formation.  We also ran another simulation with identical initial conditions but with no magnetic field for comparison.
In the simulation with magnetic fields we use an ideal MHD treatment in which the equations are solved using the MHD HLL3R solver \citep[e.g.,][]{Waagan+11}, which is suitable for highly supersonic astrophysical flows. The heating/cooling process are implemented using the analytic fits by \citet{KI02} for the heating and cooling functions, which are based in the thermal and chemical calculations considered by \citet{Wolfire+95, KI00}. The gas self-gravity is implemented using the Tree-based technique \citep[e.g.,][]{Wunsch+18} and we also consider the gravitational interactions between sink particles (see below) and gas cells.
For further information, we refer to the reader to \cite{ZA+17}.

As we are interested in following the gravitational collapse of cores, we refine according to the Jeans criterion, resolving the local Jeans length by at least 10 grid cells in order to prevent spurious fragmentation \citep{Truelove+97}.\footnote{Note that the Truelove criterion requires the local Jeans length to be resolved by at least four cells in HD simulations.  However, in MHD ten cells are a good compromise to properly follow the magnetic field amplification by gravity-driven turbulence \citep{Federrath+11}.} When the maximum resolution is reached and the Jeans length of the collapsing gas can no longer be resolved, we introduce sink particles \citep{Federrath+10}. Note, however, that the original simulations refine only 10 levels, reaching a spatial resolution of 0.03 pc which just marginally resolves the dense molecular cores. Thus, in order to resolve the structure within cores, we re-ran the simulation (just a few tenths of Myr before sink formation) allowing the code to refine a further eight levels within a radius of $\sim 5$ pc of each sink, reaching thus an effective resolution of $\sim 25$ AUs ($1.2 \times 10^{-4}$ pc). We let each core evolve until they start to form sinks at a threshold density of $n_{\rm thr} \simeq 2.2 \times 10^9 \, {\rm cm}^{-3}$.

The initial conditions are as follows. Each stream is cylindrical, with 32 pc in radius and 112 pc in length along the x-direction, containing warm gas in thermal equilibrium with density $n=2 \, {\rm cm}^{-3}$ and temperature T=1450 K. These streams are completely contained in the numerical box of size 256 pc in the $x$-axis and 128 pc in the $y$/$z$-axes. The streams collide at a transonic velocity of $\sim 7.5 \, {\rm km~s}^{-1}$, and we impose a background turbulent velocity field (with Mach number of 0.7) in order to trigger the dynamical instabilities in the shocked layer. In the MHD simulation, the magnetic field is initially uniform along the x-direction with a strength of $3 \, \mu$G.
This corresponds to an initial mass-to-flux ratio, $\mu$, of 1.5 times the critical value ($\mu_{\rm crit}$), and thus the cloud is magnetically super-critical as a whole.  We find that the mass-to-flux ratios for the cores at the time we analyze them are roughly critical on scales of $2 \times 10^4$~AU, with the exact values depending upon the scale analyzed (because there is generally significant mass on large scales). 

\begin{figure*}
	\minipage{0.445\textwidth}
	\includegraphics[width=\linewidth]{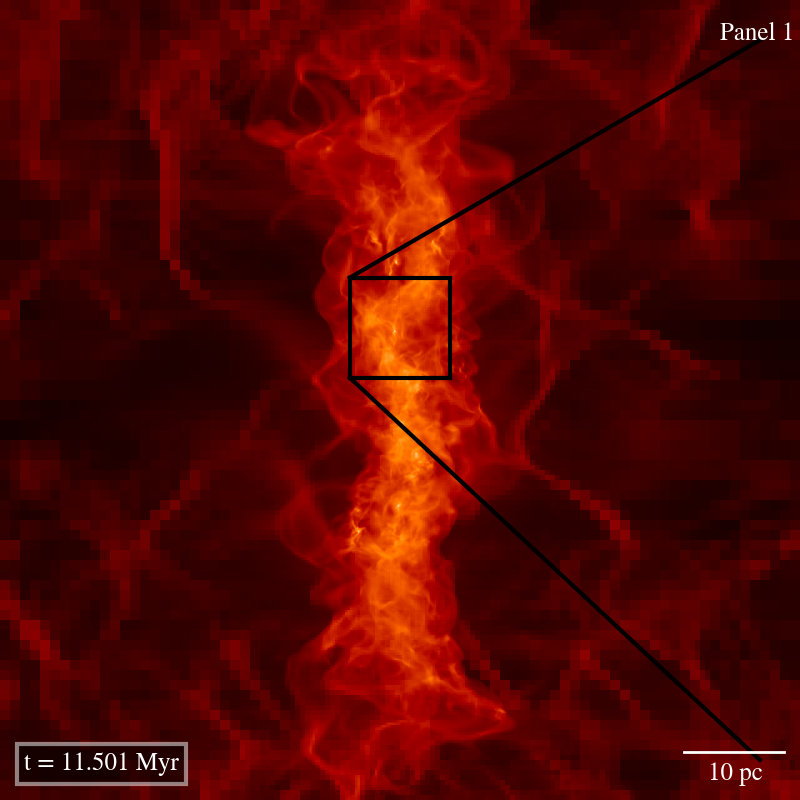}
	\endminipage\hfill
	\minipage{0.555\textwidth}
	\includegraphics[width=\linewidth]{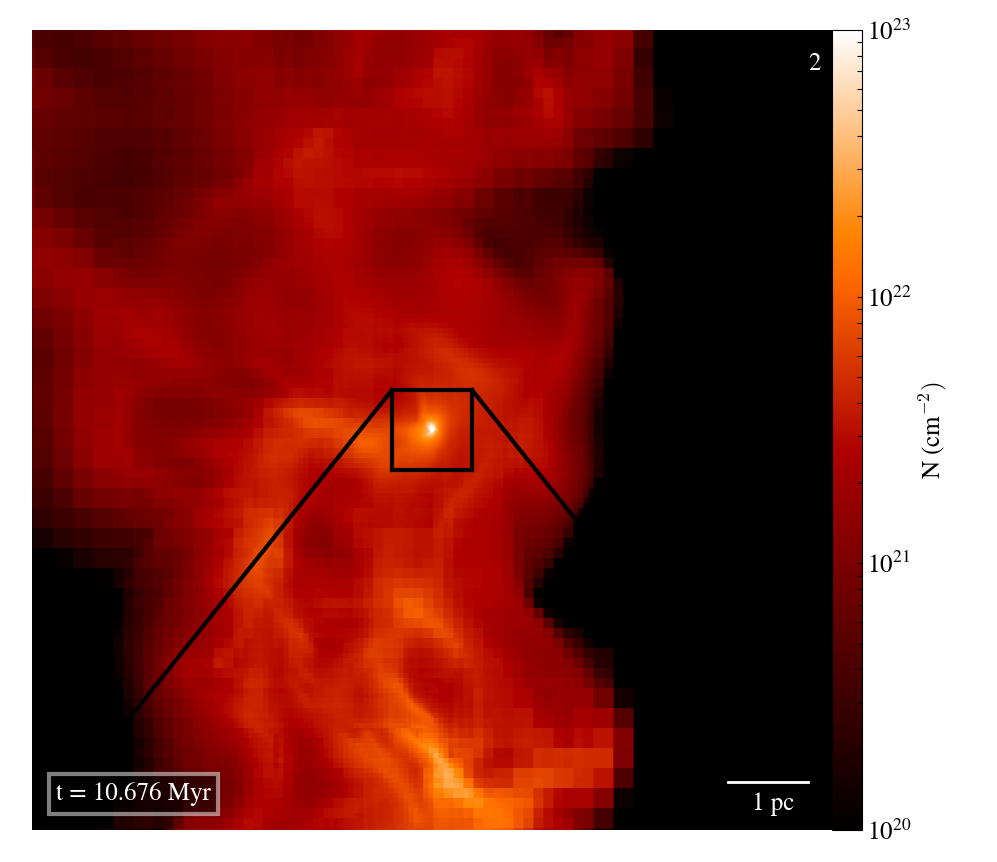}
	\endminipage\hfill \\
    	\minipage{0.445\textwidth}
	\includegraphics[width=\linewidth]{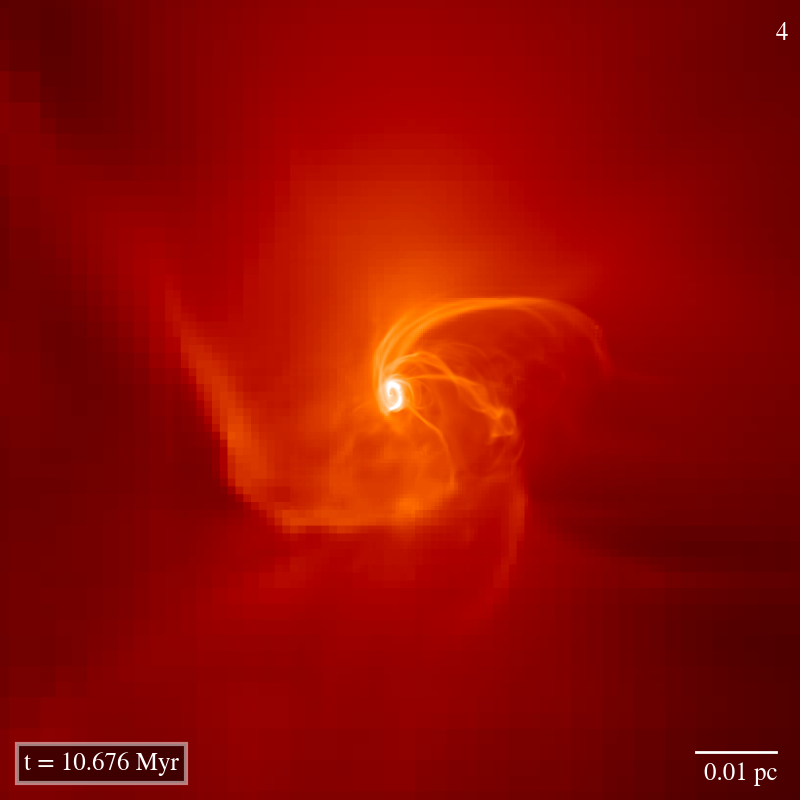}
	\endminipage\hfill
	\minipage{0.555\textwidth}
	\includegraphics[width=\linewidth]{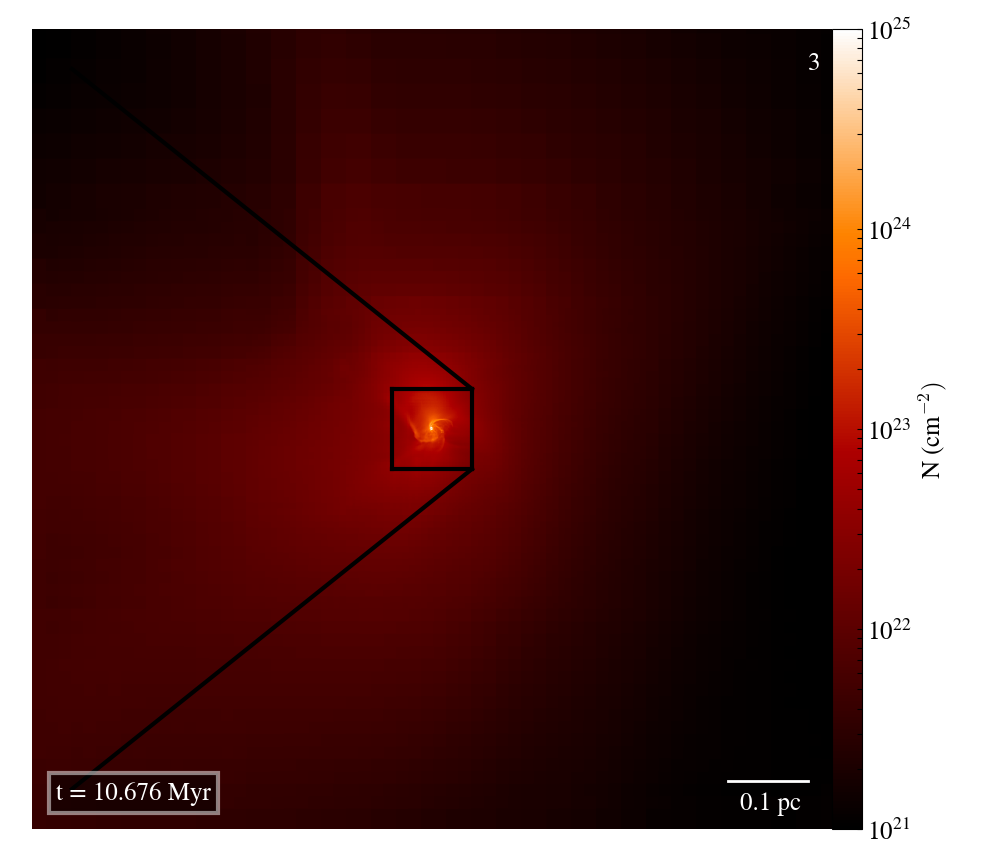}
	\endminipage\hfill
	\caption{Edge-on column density maps. Panel 1 shows the whole MC at $t \simeq 11.5$ Myr for the MHD  simulation. At this time, we restarted the simulation (for a few tenths of Myr more), zooming into the first star-forming core. Panels 2, 3, and 4 show this core in a box size of 10, 1, and 0.1 pc, respectively. Note that the color bar scale is different in the upper and lower panels. This figure was generated using the python-based {\tt yt} toolkit \citep{yt}.}
    \label{fig:cloud2core}
\end{figure*}

Panel 1 of Fig. \ref{fig:cloud2core} shows the edge-on column density ($N$) of the entire MC at $t \simeq 11.5$ Myr.\footnote{Note that the MC extends several tens of pc.} At this time we restarted the simulation for a few tenths of Myrs more, zooming, for instance, around the first star-forming core. Panels 2-4 of the same figure show projections of this core at $t \simeq 11.68$ Myr in increasingly smaller boxes.

\subsection{Post-processing}
\label{sec:post}
We made cutouts from the simulation data cubes for ten isolated cores from both the MHD and HD simulations. The masses of these cores and sinks are listed in Table \ref{table:sinks}. 
The cutouts were cubes of length 0.2, 0.1 and 0.05 pc to
analyze the properties as a function of scale.  For each
cubical cutout, the gas properties (density, velocity and temperature) were interpolated on 64$^3$ uniform grids centered on the sinks.  The gas velocities were
transformed into the rest frame of the sink, except for the low-mass sinks in \texttt{Core} 4, 8 and 16 which are moving much faster than the surrounding gas, and so we used the
rest frame velocity of the calculation.
(Presumably these sinks have been gravitationally scattered by nearby high-mass sinks.)
The temperatures were taken from the simulation
results and are in the range $\sim 9-13$~K. This ignores heating by the central protostar, but this should be quite minimal
on the scales we consider for typical protostellar luminosities.
The simulation data in the cutouts are available
upon request.

\begin{table}
\centering
\caption{Masses of sinks and enclosed gas masses $\Msun{}$
as a function of the radius in AU.}
\label{table:sinks}
\begin{tabular}{lrrrr}
\hline
Core & $M_{sink}$ & $M_{5000}$ & $M_{10000}$ & $M_{20000}$\\
\hline
1   & 2.7   &  4.0 & 7.5 & 12.8  \\
2    & 0.7  & 3.9 & 8.2 & 15.9\\
3    & 1.2  & 3.2 & 6.4  & 11.2  \\
4    & 0.3  & 4.2 & 7.3 & 10.8 \\
5    & 0.9  & 3.1 & 6.2 & 11.1\\
6    & 0.03 & 5.3 & 10.6  & 19.7  \\
7    & 1.0  & 4.1 & 8.5  & 19.5\\
8    & 0.2  & 4.2 & 8.3 & 14.8   \\
9    & 0.7  & 3.1 &  6.0  & 10.5\\
16   & 0.2  & 3.3 & 6.1 & 10.1 \\
nonB 1   & 0.046 &  4.6 & 8.4 & 13.9  \\
nonB 2    & 0.49 & 4.9 & 8.8 & 13.8\\
nonB 3    & 0.50 & 3.3 & 7.5  & 13.9  \\
nonB 4    & 4.9 x $10^{-4}$ & 3.6 & 7.0 & 12.5 \\
nonB 5    & 5.9 x $10^{-3}$ & 3.4 & 6.6 & 11.4\\
nonB 6    & 0.54  & 3.6 & 7.0  & 13.6  \\
nonB 7    &  0.47 & 2.2 & 5.1  & 12.2\\
nonB 8    & 2.7 x $10^{-5}$ & 5.8 & 12.9 & 23.4   \\
nonB 9    &  0.55 & 3.9 &  7.1  & 11.9\\
nonB 10    & 0.088 & 4.9 &  10.1  & 15.9\\
\hline
\end{tabular}
\end{table}


These cutouts were then used as input to the LIME \citep{brinch10} radiative transfer code to produce channel maps of the \ce{NH_3} (1,1) inversion lines in three orthogonal
projections. 
The velocity resolution between neighboring channels was set to be 0.1 km $s^{-1}$. The channel maps were made in three orthogonal projections denoted $XY$, $XZ$ and $YZ$, on the four different scales of the cutouts.

We do not include chemical modeling in
our analysis.  Initially, we set
the abundance of \ce{NH_3} relative to \ce{H_2}  to a uniform and relatively high value of $8 \times 10^{-8}$ \citep{aikawa99}.
We also calculated emission maps using a lower abundance for \ce{NH_3} of $5 \times 10^{-9}$; this
did not change the results of the velocity
analysis significantly.
In addition, we examined the effect of setting the ammonia
abundances to zero at densities
below $10^4  \ppcc$, as
studies of nearby cores indicate that bright \ce{NH_3} emission generally occurs at higher densities \citep[e.g.,][]{myers83}.  Again, this
made no significant difference to the results,
especially because
all of the grid cells within a radius of 5000 AU of the sinks have densities above $10^4  \ppcc$.

For each cutout, we used the Python package \texttt{spectral\_cube} \citep{spectralcube} to make moment maps (intensity maps, intensity-weighted velocity maps and line-width maps, or moment 0, 1 and 2 respectively) and the Python package \texttt{pvextractor}  \citep{pvextractor} to make position-velocity (p-v) diagrams. The p-v diagrams were made by slicing along the direction of the velocity gradients measured on the various scales (see \S \ref{sec:vgrad}).

The radiative transfer solutions have convergence
problems for some cores, such as \texttt{Core} 2. We believe that this is due to the limitation of LIME's linear interpolation scheme for the velocity field. Problems arise when the velocity differences between neighbor cells are greater than the turbulent velocity dispersion (Brinch, personal communication). Smoothing our complex supersonic velocity fields with a 3-D Gaussian kernel did not help solve the problem. 
The only method we found to promote satisfactory convergence in the problem cases with the current version of LIME and
our computational resources was to increase the turbulent velocity "$doppler$" parameter from 0.2  $\kms$ to 0.8 $\kms$. 
To see whether this smoothing has a
significant effect on our results,
we also generated moment maps with simple volume density-weighted ``emission'' to be analyzed in the
same way as the maps generated with
LIME.  As discussed in the results section,
while there are significant differences in
some projections for a few cores, the overall
velocity gradient results did not change.

\subsection{Velocity gradient analysis}
\label{sec:vgrad}

We fit for the "observed" velocity gradient using the method that \cite{goodman93} adopted to measure ammonia observations of cores. The method involves 3-D least square fitting of the moment 1 map with the following equation:
\begin{equation}
    v_{LSR} = v_0 + a\Delta x + b\Delta y\,,
\end{equation}
where $v_{LSR}$ is the velocity at each pixel, $v_0$ is the average velocity over the map (relative to
the sink),
$a$ and $b$ are the fitting parameters, and $\Delta x$ and $\Delta y$ are the displacements on x-axis and y-axis, where x and y are the displacements perpendicular to the particular line of sight in the map. 
The mean angular velocity is then 
\begin{equation}
    \Omega = \sqrt{a^2+b^2}
\end{equation}
in units of $\kmspc$. The position angle of the velocity gradient is then given by
\begin{equation}
    \theta = tan^{-1}(\frac{b}{a})\,.
\end{equation}

The $\chi^2$ statistic was used to optimize $a$ and $b$,
\begin{equation}
 \chi^2(a,b) = \sum_i \left({v_{LSR} - f(x_i,y_i|a,b) \over \sigma_{v_{LSR},i}}\right)^2.
\label{equ:fitting}
\end{equation}
The uncertainty of the velocity in the fitting was estimated following at each pixel \cite{uncertainty};
\begin{equation}
\sigma_{v_{LSR}}=1.15(\frac{\sigma_T}{T})(\delta_v\Delta v)^{1/2}\,
\end{equation}
where (${T}/{\sigma_T}$) can be treated as the signal to noise ratio {\bf (SNR)}. We assume SNR $=5$ and constant over the moment 1 map. The parameter $\delta_v$ is the linewidth (moment 2) map, and $\Delta v$ is the size of the velocity bin, set to 0.1 $\kms$. 
A least square fitting method ({\bf Eq.} \ref{equ:fitting}) was used to find the best fit-values of $a$ and $b$. 

The velocity gradients were calculated on four different scales: 2500 AU, 5000 AU (0.025 pc), 10000 AU (0.05 pc) and 20000 AU (0.1 pc) in radius (half of the box length).  We calculated \ce{NH_3} maps for all the four scales and the three main projections.

\subsection{Specific angular momenta}

The velocity gradient measurements were
used to make estimates of the specific angular momentum with the assumption of solid body rotation and spherically-symmetric density profile, characterized by parameter $p$,
\begin{equation}
    j_{2D} = pR^2\Omega ,
\end{equation}
where $j_{2D}$ denotes the observationally-inferred specific angular momentum, $R$ is the radius of the map that is analyzed, and $p$ is related to $\alpha$ in the radial density profile, which can be calculated by integrating the moment of inertia of a sphere with density profile characterized by $\alpha$. If the density profile is a power-law,
\begin{equation}
\rho \propto r^{-\alpha}\,,
\end{equation}
the relation between $p$ and $\alpha$ is 
\begin{equation}
    p = \frac{2(3-\alpha)}{3(5-\alpha)}\,.
\end{equation}
For a singular isothermal sphere, with
$\alpha$ = 2, $p = 2/9$, while for
a uniform density sphere $p = 2/5$.

\begin{figure*}
	\minipage{0.9\textwidth}
	\includegraphics[width=\linewidth]{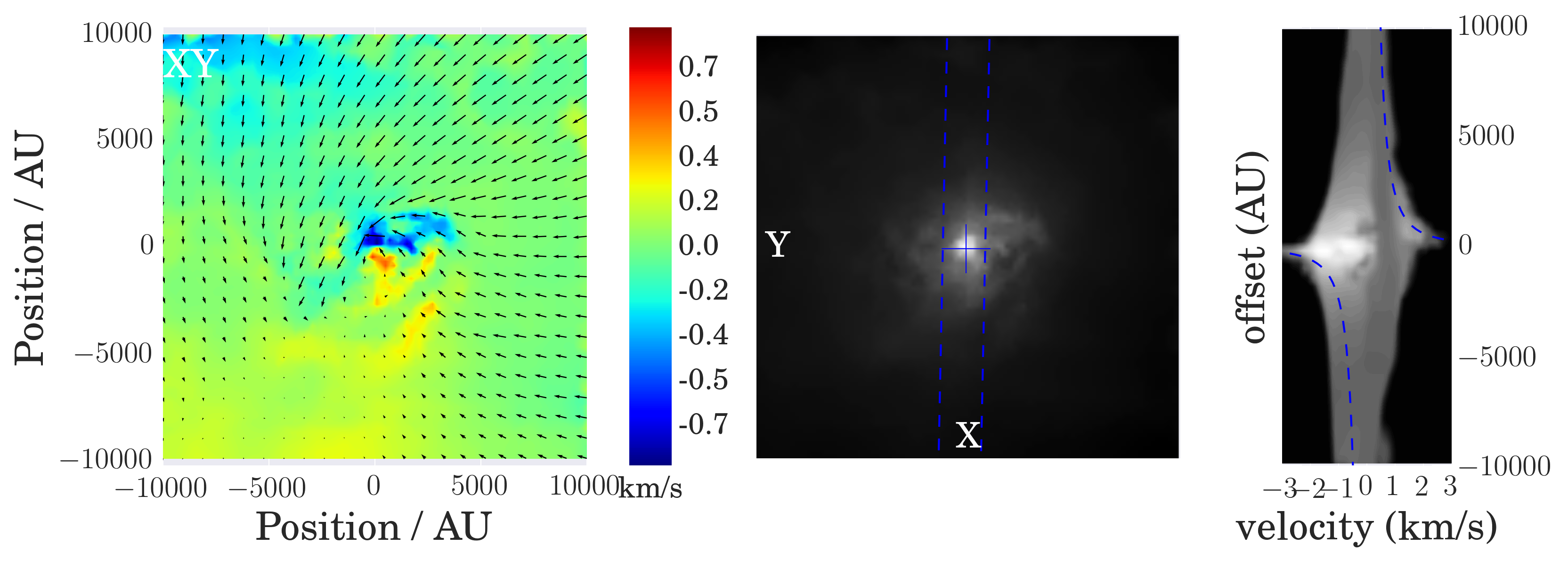}
	\includegraphics[width=\linewidth]{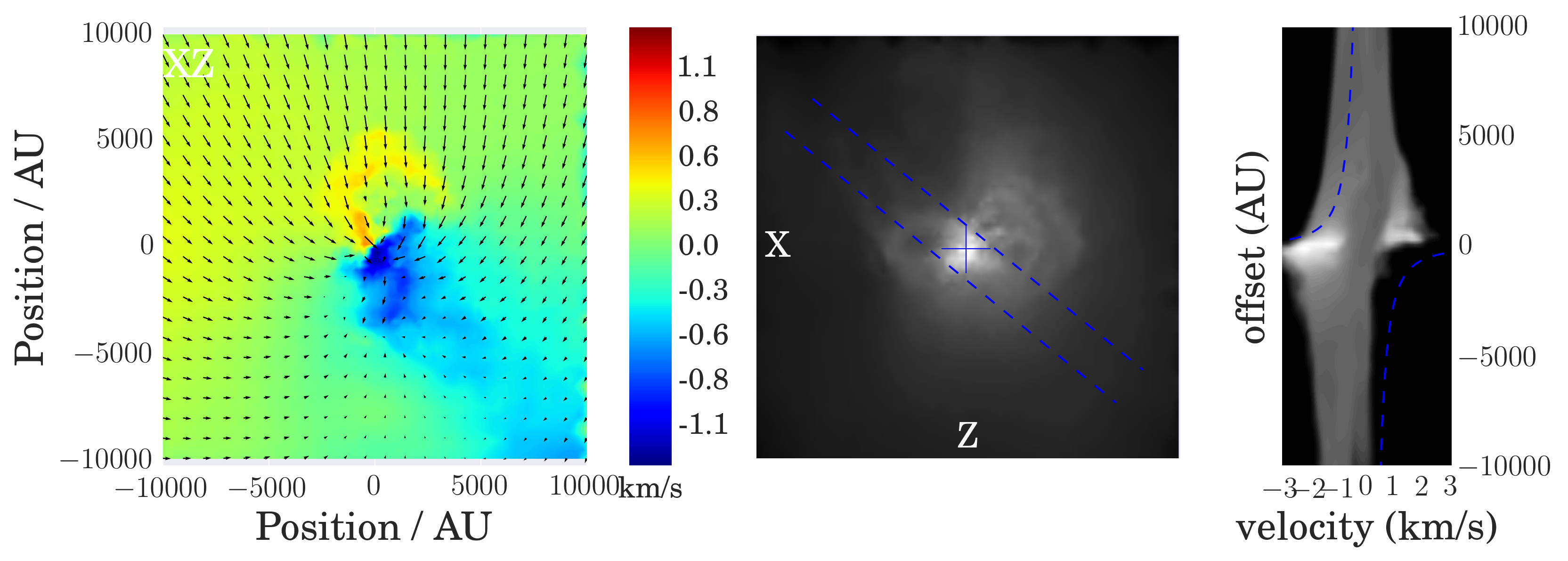}
	\includegraphics[width=\linewidth]{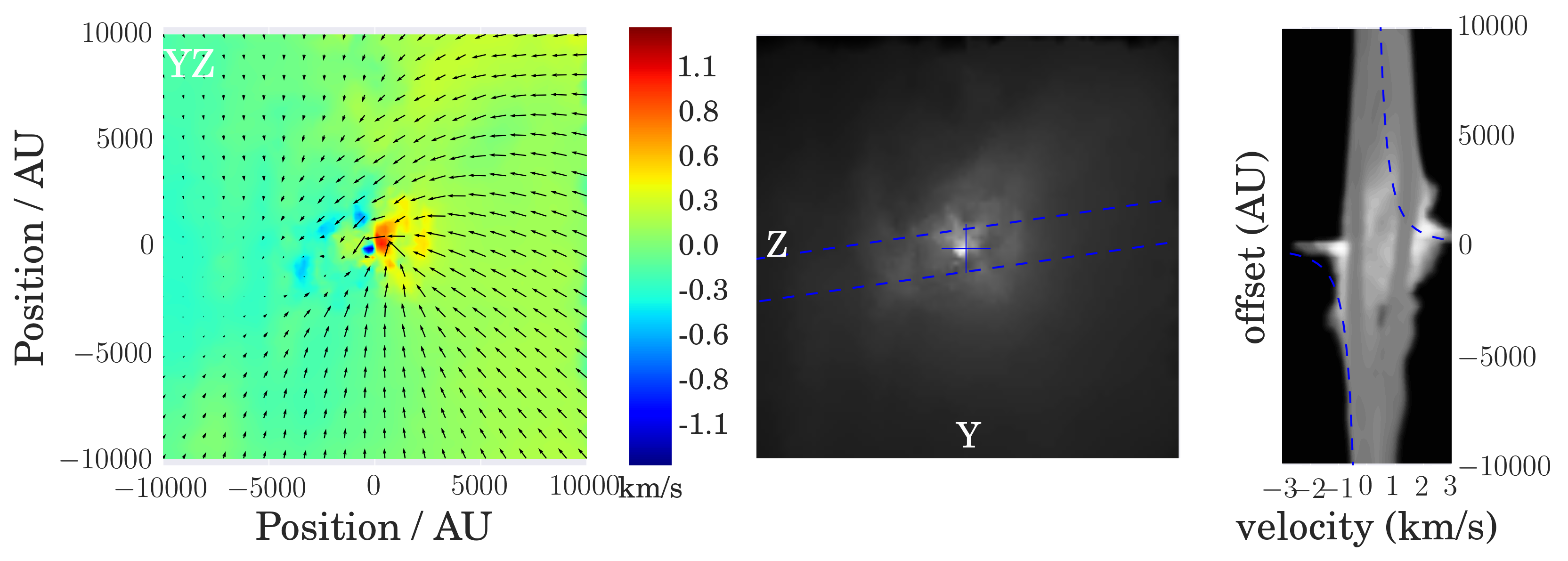}
	\endminipage\hfill
	\caption{Moment maps and p-v diagram for \texttt{Core} 1 on a 0.1 pc 
	scale. The nine panels show the comparable data for the simulation core in three projections ($XY$, $XZ$ and $YZ$), with the left panels showing the moment 1 radial velocities in color (and velocity vectors in the plane of the sky), the center panels showing the moment 0 \ce{NH_3} map with dashed blue lines showing directions of fitted velocity gradient. The right panels are position-velocity map taken along the direction with certain width found from the velocity gradient fitting, indicated by the dashed blue lines. The dotted curves in the right panels are the Keplerian rotation velocity given the sink mass. 
	For $XY$ projection, the positive offset is from the upper region of the image, for $XZ$ projection, the positive offset is from the lower right region and for $YZ$ projection, the positive offset is from the right region of the image.}
    \label{fig:core1_20000}
\end{figure*}

The true (mass-averaged) specific angular momenta were calculated numerically within cutouts at the
different scales (cubic boxes) as
\begin{equation}
    j_{3D} = |{\bf J}/M| = \frac{\Sigma({\textbf{r}}_i\times {\textbf{v}}_i)m_i}{\Sigma{m_i}}\,,
    \label{eq:truej3d}
\end{equation}
where ${\textbf{r}}_i$ is the distance between the sink and the gas cell and $m_i$ and  ${\textbf{v}}_i$ are the mass and velocity at each cell.

It is instructive to compare the angular momenta of our cores with
some estimate of the maximum values possible while still remaining
gravitationally-bound.  There is no single way of defining such
maxima for objects in non-uniform rotation with arbitrary and
complex density and velocity distributions, as is the case here.
As a benchmark we adopt the results for the Mestel disk \citep{mestel63}.
This choice is motivated by two factors:
first, a maximally-rotating cloud should be highly flattened; 
and second, the mass of the Mestel disk grows linearly with the radius,
roughly consistent with the behavior of our cores (\S \ref{sec:results}).
Thus we use the simple relation (for the formally infinite disk)
$v^2 = G M(r)/r$, where $M(r)$ is the mass interior to $r$; thus 
\begin{equation}
j_{max} = 0.5 (G M(r) r)^{1/2}\,,
\label{eq:j_max}
\end{equation}
where we use $M(r) = M_s + M_g(r)$, with $M_s$ and $M_g$ being the
sink mass and gas mass, respectively.
(The sinks have negligible angular momentum.)

\section{Results} \label{sec:results}

\subsection{Synthetic observations}

Fig. \ref{fig:core1_20000} shows a typical example of the projected 
synthetic observation with \ce{NH_3} line emission (\texttt{Core} 1 on the 0.1 pc / 20000 AU scale in diameter). The density-weighted velocity vectors in the plane of the sky are overplotted on the moment 1 maps.  
The sink (protostar) is located at (0,0,0) and denoted by a cross in the moment 0 maps.  The dashed lines indicate regions where the p-v diagrams were sliced, with the directions given by the velocity gradients.

The dotted lines in p-v diagrams for each orthogonal projection indicate the Keplerian rotation curve given the sink mass. While these curves qualitatively resemble the shapes in the p-v diagrams, they underestimate the observed velocity range, not only due to projection effects but because the envelopes contain significant mass beyond that of the central sink.

\begin{figure*}
    \centering
    \includegraphics[width=\linewidth]{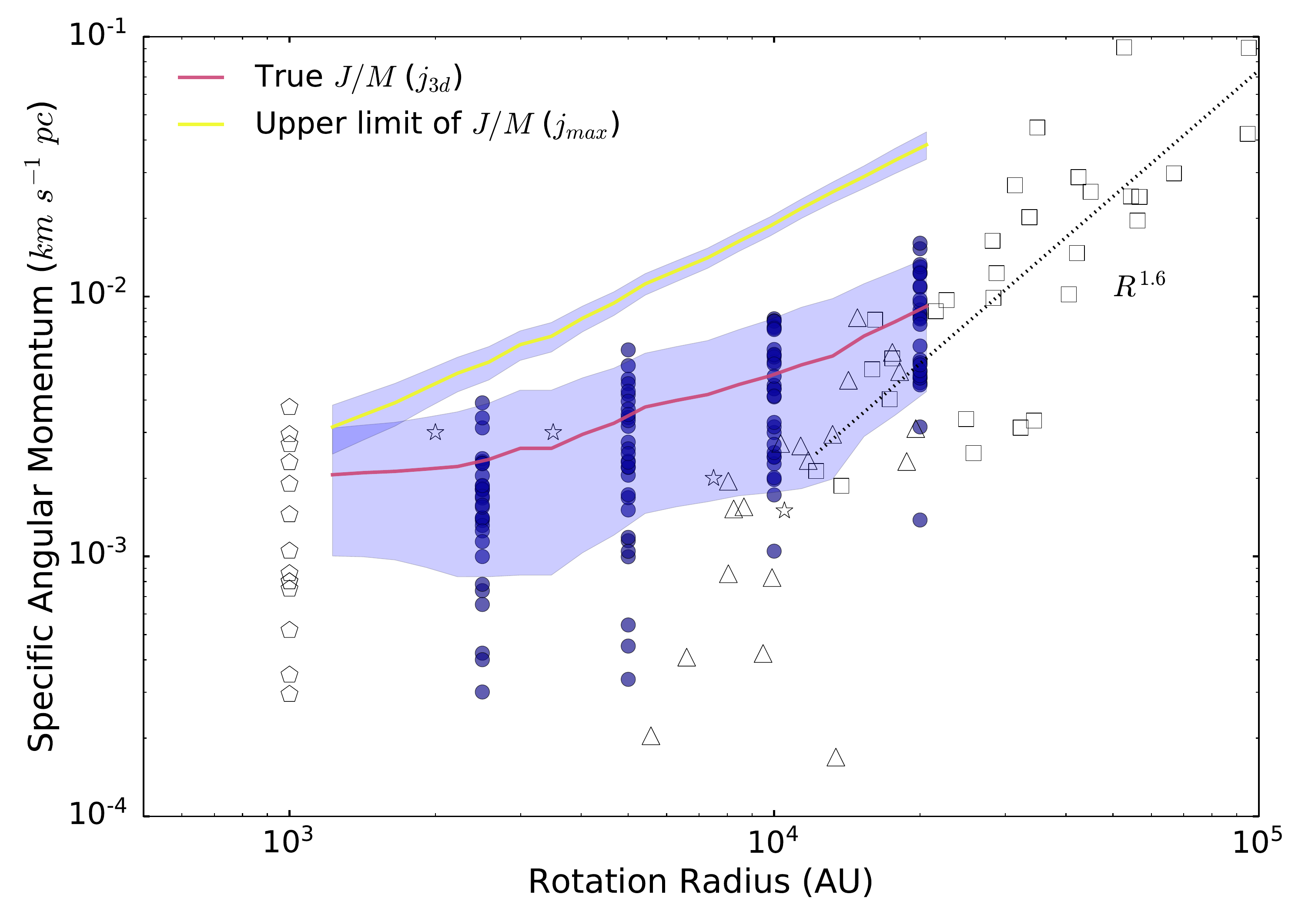}
    
    \caption{Specific angular momentum (J/M) as a function of radius for our ten MHD cores and a sample of observations. The circles are $j_{2D}$ derived from observing and analyzing the simulation in the same way as was done for the real cores. The lower solid (red) curve is the median of $j_{3D}$ for our ten cores and the light shaded area indicates $\pm  1 \sigma$ (linear). The yellow curve is our estimate of the maximum $j$ consistent with gravitational binding (equation \ref{eq:j_max}) with the darker shaded area the $1 \sigma$ dispersion. The squares are observations of $\ce{NH_3}$ dense cores \citep{goodman93}; the stars are measurements for IRAM 04191 at four different scales (\citealt{Belloche02, Belloche04}), the triangles are \ce{N_2H^+} measurements in \citep{caselli02}, and the pentagons are C$^{18}$O measurements of cores on 1000 AU scales from \citep{yen15}. The dotted line indicates the $r^{1.6}$ dependence of $j$ estimated by \protect\cite{goodman93}.}
    \label{fig:j2d_j3d}
    
\end{figure*}

\subsection{Observationally-inferred vs. true specific angular momenta}

\begin{figure}
    \centering
    \includegraphics[width=\linewidth]{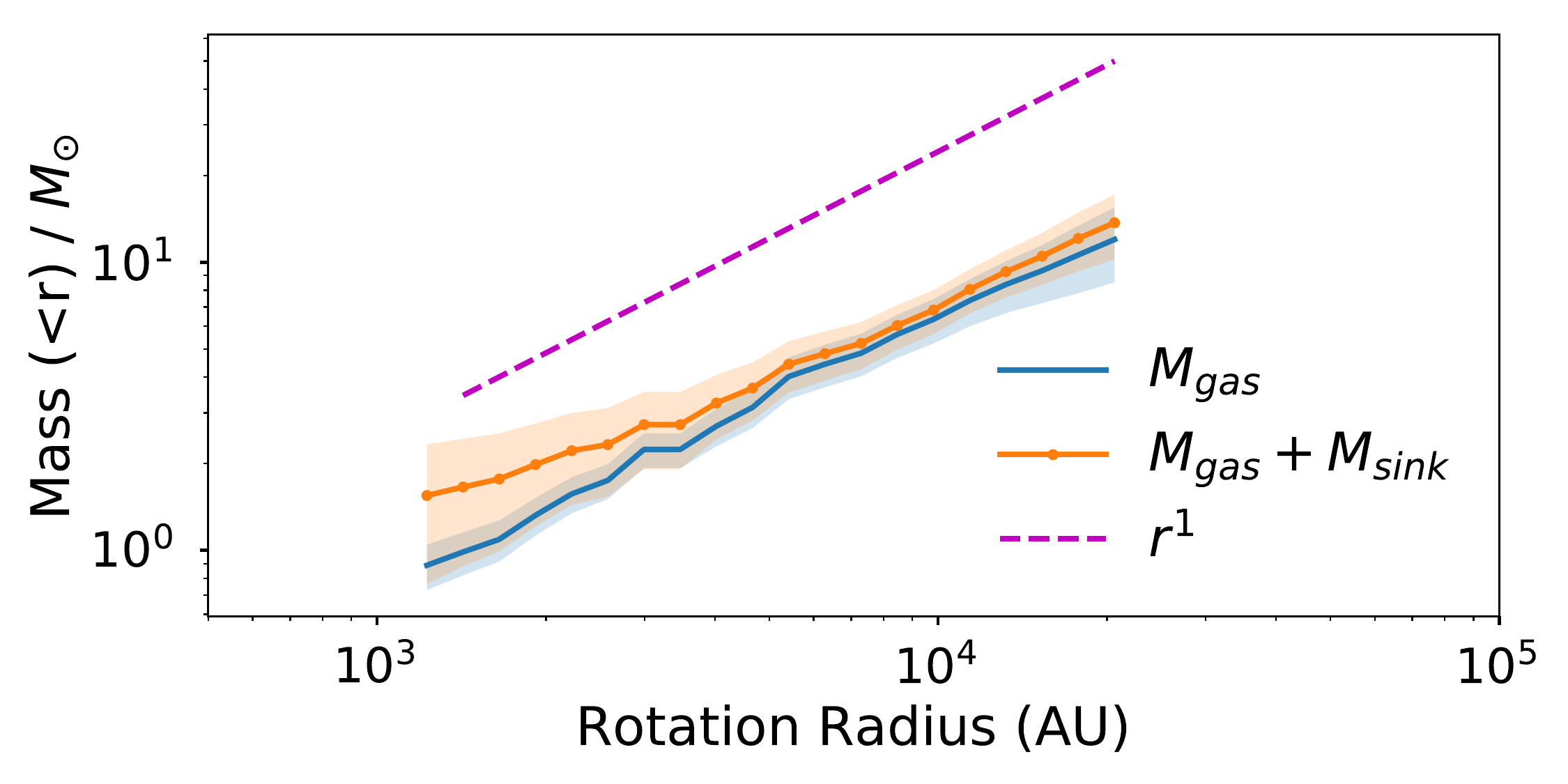}
    \caption{ Cumulative mass of ten MHD cores. The solid line indicates the median of cumulative gas mass for ten cores and the dot-solid lines indicates the median of cumulative gas mass plus the sink mass. 
    The general trend is nearly $M(r) \propto r$, except at small radii where the sink mass dominates.}
    \label{fig:cum_mass}
\end{figure}

Figure \ref{fig:j2d_j3d} shows results for the high \ce{NH_3} abundance case.  The filled
circles are the estimated angular specific
angular momenta for each projection of the ten
cores adopting $p$ = 2/5.  In most cases there
is reasonable agreement with the median $j_{3d}$ angular momenta of the cores (red curve) and
the range is also fairly consistent with the
one-sigma dispersion (shaded area).  There
are a few cases where the observations strongly
underestimate the median $j_{3d}$ angular momentum, and these
are due to projection effects (e.g., the line of sight is close to the orientation of the angular momentum vector{\bf )}.
The true specific angular momenta fall well below our simple
upper limit estimate (yellow curve and dark shaded area), except
at the smallest scales where rotational support is most important.
At scales beyond 3000 AU the slope of the upper limit is
close to unity.  This is consistent with the
variation of enclosed mass with radius; as shown in in Fig.\ref{fig:cum_mass}, 
 $M(r) \approx r$, so that the expected upper limit $ j _{max} \propto (G M r)^{1/2} \propto r$.

 Both the median $j_{2d}$ and $j_{3d}$ values are results are consistent with the results of \cite{goodman93} at 20,000 AU,
as well as those of \cite{yen15} on 1000 AU
scales. but lie above the IRAM 04191 results of \cite{Belloche02,Belloche04} and the N$_2$H$^+$ measurements of \cite{caselli02}, although some of the objects in the \citeauthor{caselli02}
sample
 are starless cores which are not directly comparable to our cores with sinks.
 
 Note that the radial dependence of $j_{3d}$ in Figure \ref{fig:j2d_j3d} demonstrates that $\Omega(r)$ is not constant but declines with increasing radius.  Nevertheless, the velocity gradient method, which assumes constant angular velocity, can still yield a reasonable estimate of the actual angular momentum in most cases.  This may be due to the total angular momentum being dominated by the motion on the largest scales.

\begin{figure}
	\minipage{0.45\textwidth}
	\includegraphics[width=\linewidth]{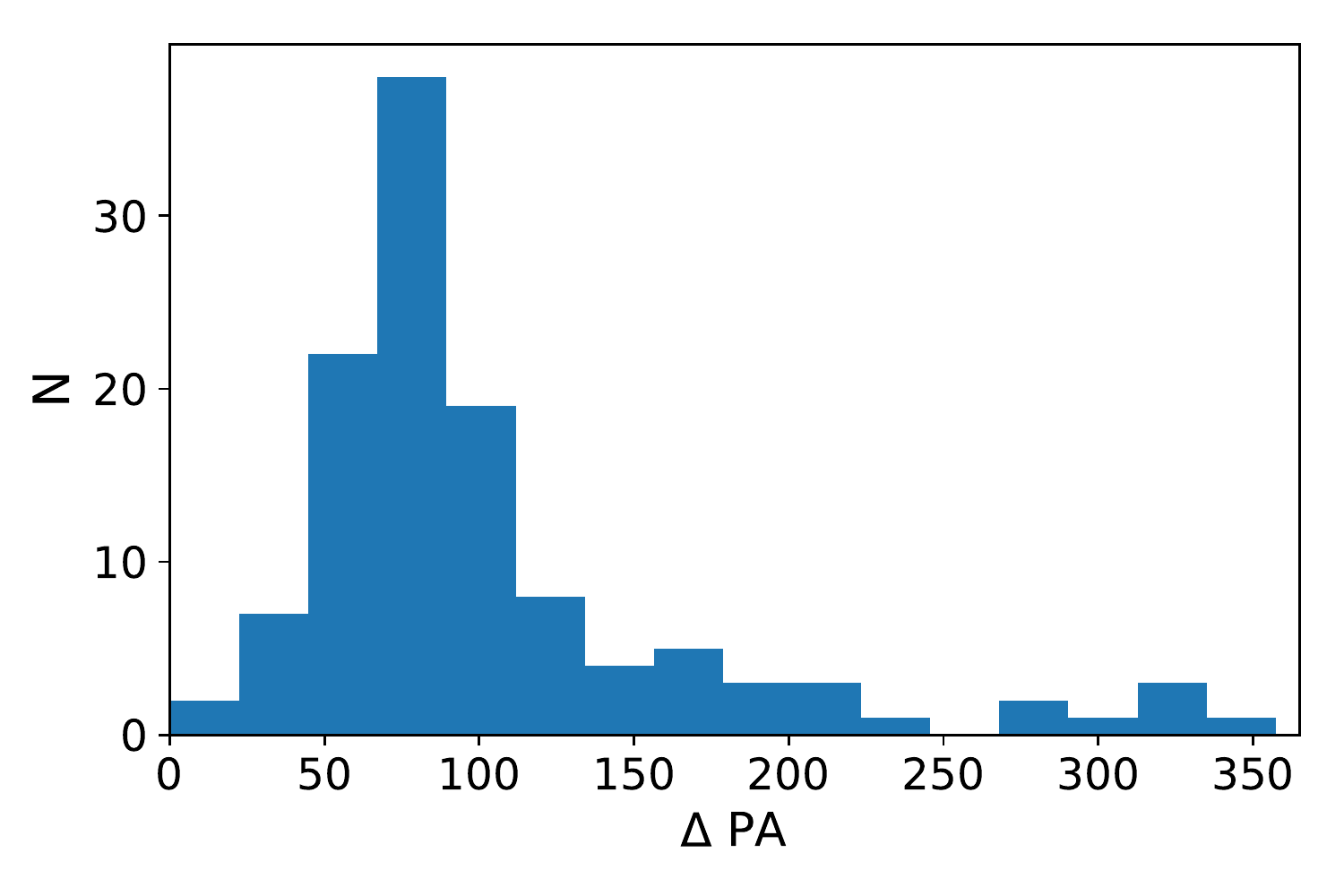}
	\endminipage\hfill
	\caption{The distribution of the angle between the velocity gradient and the projected $j_{3D}$. The difference in position angle peaks at 90$^\circ$, as would be the case for perfect alignment; about 25\% of the cores show differences of more than 45$^\circ$, with about 10\% of the cores deviating by more than 90$^\circ$. This is due to the complexity of the velocity field.}
	
    \label{fig:angle}
\end{figure}

It is also of some interest to consider the reliability of observed position angles of rotation.
In Fig. \ref{fig:angle} we show a histogram of the difference in position angle between the projected velocity gradient and the projected direction of the true angular momentum vector, calculated for the ten MHD cores on the four different scales. 
The direction between the velocity gradient and projected $j_{3D}$ peaks at around 90$^\circ$ which would indicate complete agreement between true and inferred projected position angles, but there is a dispersion around this value of about $45^{\circ}$, and in about 10\% of the cores (those with $PA > 270^{\circ}$ the sense of the rotation is opposite that of the true direction (using a right-hand rule).

\subsection{Sensitivity to radiative transfer}

\begin{figure}
	\minipage{0.45\textwidth}
	\includegraphics[width=\linewidth]{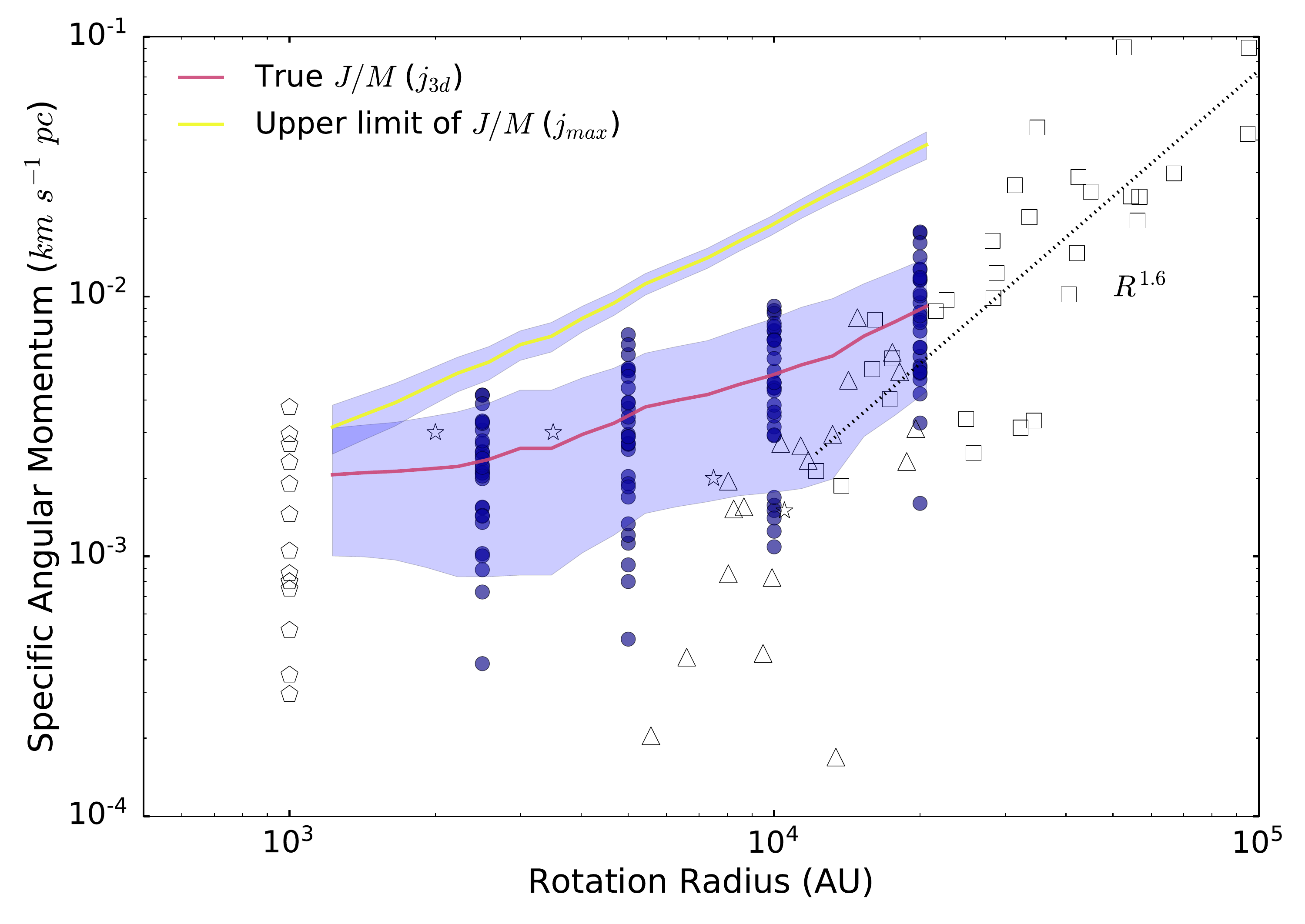}
	\endminipage\hfill
	\caption{Comparison of the true angular momenta with the velocity gradient estimates, now assuming simple density weighting to construct the moment 1 map.  The results are very similar to the radiative transfer results (cf. Fig. \ref{fig:j2d_j3d})}.

    \label{fig:non_rad}
\end{figure}

\begin{figure}
	\minipage{0.45\textwidth}
	\includegraphics[width=\linewidth]{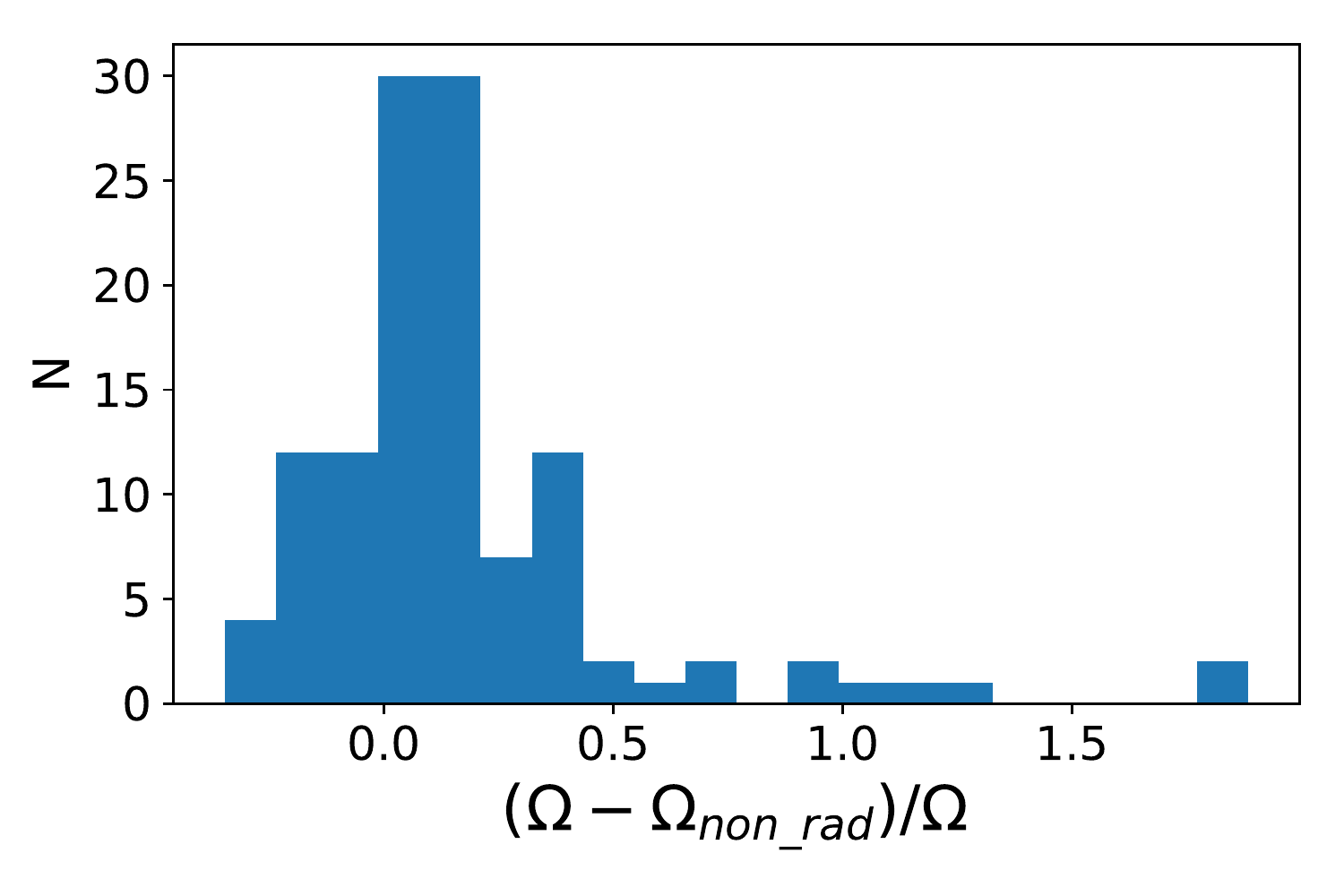}
	\includegraphics[width=\linewidth]{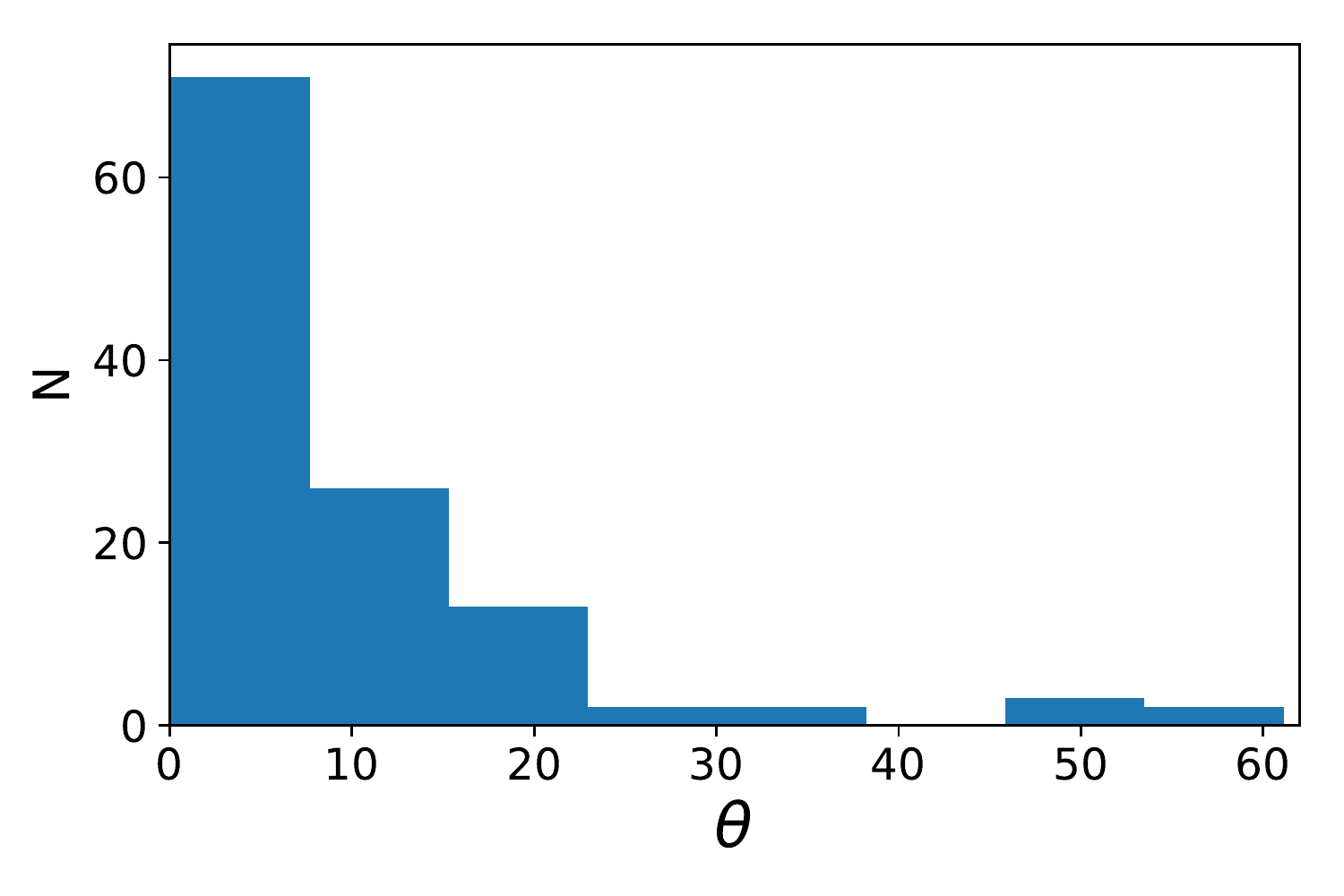}
	\endminipage\hfill
	\caption{Upper: The distribution of relative difference between the all velocity gradient results (for ten cores in MHD run on four scales and three projections) after radiative transfer and without radiative transfer. The difference of 96\% of projections is within a factor of two.  Lower: the distribution of the difference between position angle of the velocity gradient after radiative transfer and without radiative transfer for the same data above. Most (about 90\%) of the projections differ by less than 30 degrees.}
    \label{fig:wwo_rad}
\end{figure}

As noted in \S \ref{sec:post}, in some cases we had to increase the turbulent velocity to unrealistically high values in order to obtain convergence for the \ce{NH_3} radiative transfer solutions.  To test the sensitivity of our findings to the transfer approximations, we constructed moment maps that were simply density-weighted.  
As shown in Figure \ref{fig:non_rad}, there is little difference with and without radiative transfer solutions, indicating that the main
findings are robust.  There are of course significant differences in individual cases, as shown in Figure \ref{fig:wwo_rad}; even so, the estimated velocity gradients generally agree within a factor of two, which is comparable to the spread in estimated angular momentum in either case.  Again, there are extended tails in the gradient and position angle differences, emphasizing that the  observation of any single core in a particular projection is subject to significant uncertainties due to the complex geometry of
the velocity field.

\begin{figure}
	\minipage{0.45\textwidth}
	\includegraphics[width=\linewidth]{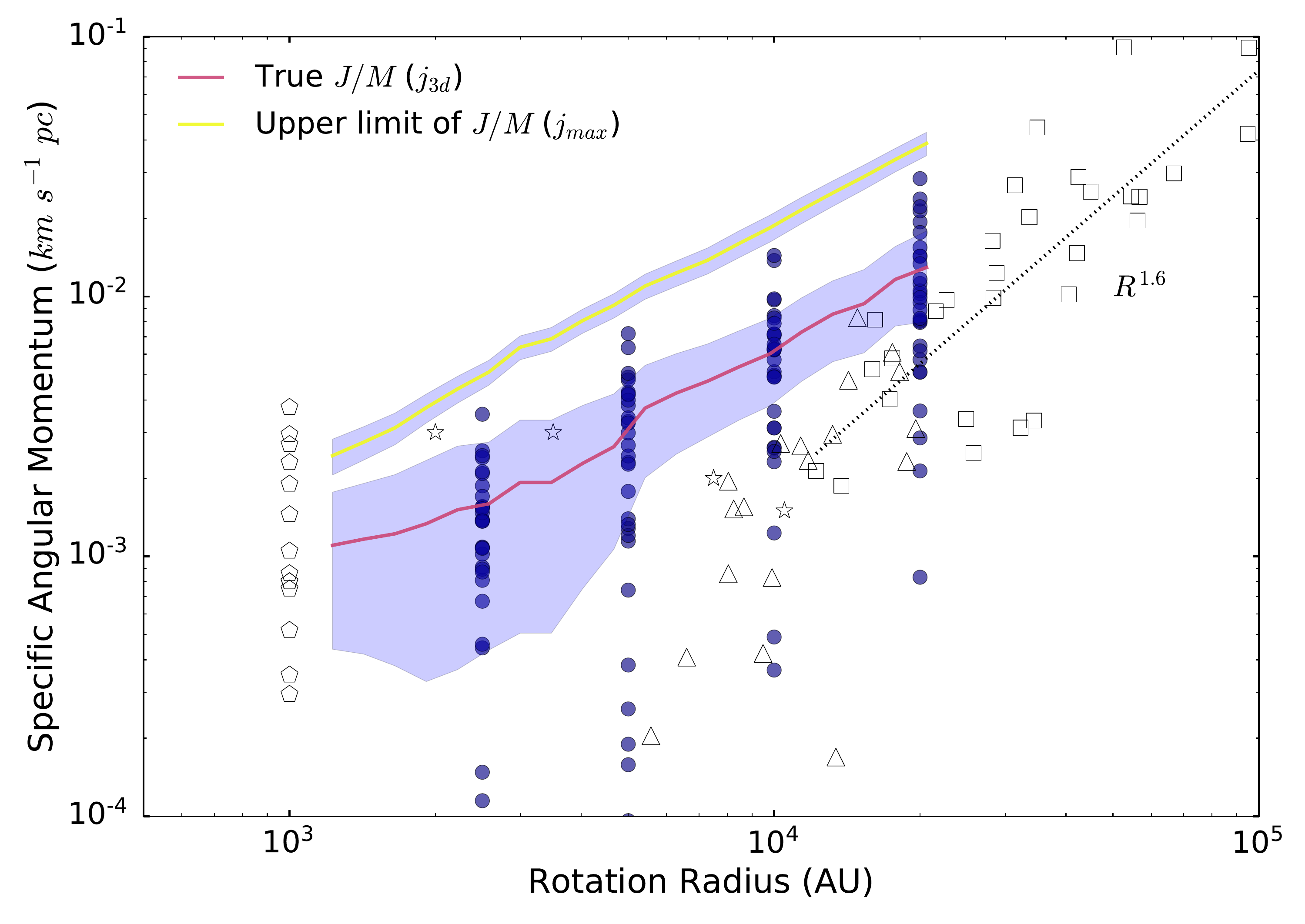}
	\endminipage\hfill
	\caption{The comparison of $j_2d$ and $j_{3d}$ for non-magnetic field runs.
	While the behavior is similar to that of the MHD run, the non-magnetic case
	exhibits some very small values in projection, mostly for the very small masses of some of the sinks in the non-magnetic run (Table 1).}
    \label{fig:nonB_j23}
\end{figure}


\subsection{Non-magnetic simulation}

We have also performed the same analysis for ten cores for the non-magnetic field simulation with the same size cutouts of the cores.  The results shown in  Figure \ref{fig:nonB_j23} again show very
similar behavior to that seen for the magnetic cores, except that the dispersions are larger and there are more projections with very small apparent angular momenta. This seems to be
due to the sinks with very small masses, resulting
from their later formation \citep{ZA+17}.  In essence, these very low-mass objects are the equivalent of ``starless cores'' and
so are qualitatively different than the other cores with
significant central gravitating masses.

\begin{figure}
	\minipage{0.45\textwidth}
	\includegraphics[width=\linewidth]{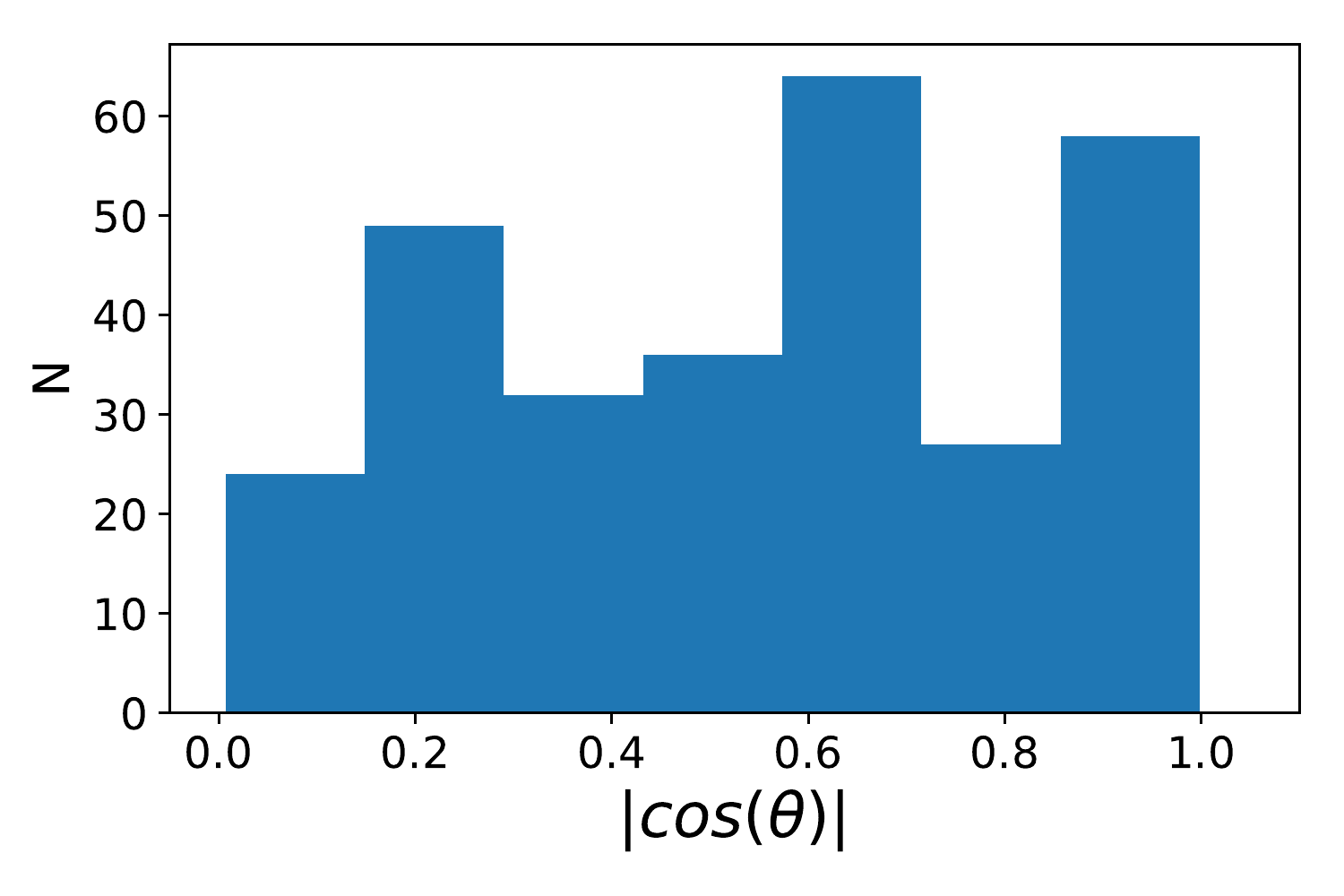}
	\endminipage\hfill
	\caption{Histogram of $cos\theta$, where $\theta$ is the angle between $j_{3d}$ and x-axis (the direction of initial B-field). The orientation of the $j_{3d}$ is consistent with a random distribution.}
    \label{fig:J_initB}
\end{figure}

 Given the similarity of the magnetic and non-magnetic runs, we examine whether the presence of the magnetic field affects the directions of the core angular momenta.  Figure 
 Fig. \ref{fig:J_initB} shows a histogram of the angle between the true angular momentum vectors and the x-axis - which is the direction of the initial magnetic field.  The differences are consistent with a random orientation, indicating that the magnetic field is not playing a major role in ordering the spins of the cores. 

\section{Discussion} \label{sec:discussion}

Our results differ from the findings of the simulations of \cite{dib10} and the driven simulations of \cite{offner08}, who found that the $j_{2D}$ is generally about 10 times higher than $j_{3D}$. We argue that
this is probably due to the more
complex velocity fields
in those simulations than in ours,
in which we allow turbulence to decay
for a considerable period of time.
This suggestion is consistent with
the findings of
\cite{offner08}, who found that their
simulation with decaying turbulence
led to a much smaller bias in
observationally-estimated angular
momentum.

The reason
for our disagreement with the results of \cite{dib10}
is somewhat less clear, as those simulations did not include continued driving; however, we suggest that
their simulations probably did have more small-scale turbulence.
The Dib \etal results refer to 
a relatively early stage of the collapse, at $\sim 0.4$ free-fall times; and it seems plausible that,
although the input turbulent velocity
field was decaying, it still dominated
the small scale motions.  In support
of this interpretation, we note that Dib \etal found that the Mach number of the turbulence increases at later times due to collapse in the cores, so
we suggest that this is signature
of gravitationally-driven turbulence rather than a remnant of the initial
imposed velocity field.  In our
simulations, which address later stages of collapse with the buildup
of significant amounts of mass in the sinks, it is clear that gravity is the
main driver of the supersonic motions, which apparently results in a supersonic velocity field that, while complex, does not strongly bias observational estimates.
In any case, for the stage of evolution considered here, we find no support for the suggestion of Dib \etal that the global gradient method with the assumption of uniform rotation generally will result in large overestimates of specific angular momentum.

Using an appropriate value of of
the ``fudge factor'' $p$, we find that our ``observational'' estimates of the specific angular momentum fall below the maximum set by gravitational binding.  One might think that if observational estimates generally indicated values that were too high by an order of magnitude, that would show up as an apparent finding that the cores wouldn't seem to be bound.  Of course we have the advantage that our $M(r)$ values are exact rather than observationally estimated as well, but such a large general discrepancy might be noticed.

Conversely, we find that for certain projections our ``observed'' specific angular momenta can be an order of magnitude lower than the true values.  We interpret this quite simply as the result of the orientation of the line of sight to the angular momentum vector, which occasionally should be unfavorable.  This is another area in which we differ
from the Dib \etal results, where there was no case in which the two-dimensional projection underestimated the true value of $j$.

These differing results
emphasize the importance of understanding
the nature of ``turbulence'' in
molecular clouds.  In our simulations,
while there is an initial turbulent
velocity field which seeds structure,
at the later stages where we observe
our cores, complex gravitational accelerations produced by the non-uniform density fluctuations and sinks
are driving the motions.  This scenario
is consistent with the results of
\cite{seifried18}, who performed large-scale numerical simulations 
of molecular clouds created and perturbed by external supernovae.
Seifried \etal found that the initial
turbulence created during cloud formation decays rapidly as the cloud becomes more massive, concluding that
SNe are generally not able to sustain observed molecular cloud turbulence.
Our simulations therefore may be more realistic
than those with continuously injected turbulence
by an unspecified mechanism.

The motivation for simulating core formation and collapse in large-scale simulations is to allow the initial conditions - the angular momentum and the detailed velocity field - to develop naturally.  Our results and those of other simulations on molecular cloud scales may be helpful in providing more realistic initial conditions for small-scale simulations of individual cores such as the recent investigations by \cite{matsumoto17} and \cite{lewis18}.  For example, \cite{matsumoto17} suggest that $j \propto r^{3/2}$ is to be expected from turbulence, which is considerably different from what we find in our simulations.
More large-scale simulations are needed with a wider variety of global initial conditions to explore this further.

\section{Conclusions} \label{sec:conclusion}

We have analyzed the results of numerical simulations of star formation with and without magnetic fields with decaying turbulence, with and without radiative transfer
post-processing.  Using the same methods as typically used to obsevationally estimate specific angular momenta $j$ for protostellar cores,
we find that even though the assumption in these methods of uniform rotation and simple power-law density distributions are not correct in detail, the methods can still serve to produce estimates of $j$ that are within a factor of two to three of the real values.  These results are in contrast with other findings in cases where the turbulence is continually driven by an unspecified mechanism.

Our simulations do not include stellar feedback.
Outflows from low-mass protostars pose a particular challenge to distinguish expansion from undisturbed core velocity fields \citep[e.g.,][]{tobin11}, which can dominate uncertainties in estimating core angular momenta. Simulations including schematic
bipolar outflows could help constrain these uncertainties. In any case, our results suggest that it is worthwhile and important to continue observational efforts to estimate angular momenta of protostellar cores, especially given the significance of such studies for understanding the formation of protoplanetary disks.

\section*{Acknowledgments}

We acknowledge a helpful and timely
report from an anonymous referee.
This work used computational resources and services provided by Advanced Research Computing at the University of Michigan, Ann Arbor; and IRyA-UNAM, M\'exico.
We used \texttt{matplotlib}
\cite{hunter07} and \texttt{numpy} \cite{vanderwalt11} for plotting and analysis.
This work was supported in part by the University of Michigan and 
CONACyT grant number 255295.

\bibliographystyle{mnras}
\bibliography{refs}

\bsp	
\label{lastpage}
\end{document}